# Secure solutions for Smart City Command Control Centre using AIOT

**S. Balachandar**
Research Scholar, VTU-RC, Department of MCA
CMR Institute of Technology
Bengaluru – 560 037, INDIA
aaathibala@gmail.com

**Dr. R. Chinnaiyan**
Associate Professor,
Department of Information Science & Engineering
CMR Institute of Technology
Bengaluru- 560 037 - INDIA
vijayachinns@gmail.com, chinnaiyan.r@cmrit.ac.in

**Abstract:** To build a robust secure solution for smart city's IOT network from any Cyber-attacks using Artificial Intelligence (AI). In Smart City's IOT network, data collected from different log collectors or direct sources from cloud or edge should harness the potential of AI. The smart city command and control center team will leverage these models and deploy it in different city's IOT network to help on intrusion prediction, network packet surge, potential botnet attacks from external network. Some of the vital use cases considered based on the users of command-and-control center.

***Keywords****- Artificial Intelligence, Internet of Things, Smart City, IOT Security, Smart City command and control center*

I. **INTRODUCTION**

The Internet of Things market will grow from 170 Billion devices (as on 2017) to 561 Billion devices by 2020 as reply.com [1]. It communicates a loud and clear message to both consumer IOT and Industrial IOT players in the market. It will bring more niche devices like Smart Home appliances, Smart Home Security, Digital Assistants and Home Robots from different providers. It also raises the concern on its security as per recent survey. The Artificial Intelligence (AI) growth is phenomenal since last five years, the industry research report clearly states that AI market is expected grow significantly between 2020 to 2030. as per Grandview paper [2]. It indicates that AI is integral part of other technologies like Big Data, IOT, Cyber Security, Cyber-physical systems. In this paper we will be discussing how IOT will harness the potential of AI in the IOT Command center use cases for a Smart City platform [3]. The data collected from either IOT Edge devices, IOT Aggregators or Gateways or cloud platform will be used by the AI (Deep Learning/Machine Learning models, Chatbots, Voice to Text message conversion, Smart Assistants for Support Staffs) which helps to monitor, control, and predict the IOT devices or sensors, or equipment used across the city network. The unification of AI with IOT will reap the benefits for the control room staffs or engineers to take informed decision and other activities such as "Failure Prediction", "Remote Troubleshooting", Optimized "Planned & Preventive Maintenance", "Predictive Maintenance" for the connected things from the command center.

In Section II we will elaborate the current and historical information including practical findings as well as hypothetical and methodological contributions regarding IOT Data Collection, Smart City Command Center functions and roles, Role of AI models being used in IOT Traffic and log data and what are the different models which are best suited for Predictive Maintenance and Preventive Maintenance of "Things" which are connected in the Smart City command center. In section III, we elaborate the functional needs and data flow of this approach. In section IV we provided details regarding components and how IOT and AI are integrated well within the Smart City Network and Command Center. In section V we will mention about problem relevancy with different use cases and its approaches being evaluated. We will share the experiment and results in section VI. The deployment considerations and issues are highlighted in section VII. We will present our significance of this research and share our recommendations and conclusion in Section VIII.

## II. LITERATURE REVIEW

In the "Artificial Intelligence for Securing IoT Services in Edge Computing: A Survey [4]" research journal published by "Zhanyang Xu", "Wentao Liu", "Jingwang Huang", "Chenyi Yang", "Jiawei Lu" and "Haozhe Tan" on 14 Sep 2020, they explained and detailed out the following on their survey.

1. Smart Cities is one of their application of IOT in the "Edge Computing" and the usage of IOT Edge to use to examine and troubleshoot key incidents or actions.
2. Data Privacy protection for IOT Edge devices and services with Artificial Intelligence: The need of data security (e.g., data puzzlement, encoding) methods to safeguard the data privacy while transferring the data between devices or cloud or edge network.
3. Lightweight privacy preservation using "Artificial Intelligence": They have explained two important deep learning model CNN (convolutional neural network) and deep neural network (DNN) models which are further being analyzed. CNN uses "LAYRNT" and Modified CNN Inference Module algorithms for "Privacy Preservation", both are proven with high accuracy of 91% and 92.6% however it expects high computing power to process this algorithm.
4. Device authentication, authorization management, IOT Data sharing can also be solved through "Blockchain for IOT Edge enabled devices."
5. The key messages out of this journal mentioned about "Deep Learning" will require extensive compute for such IOT security analytics use case and it may fail to uncover intruders or cyber-attacks accurately due to too many parameters (features) or not enough (complex) data points, where as the "Reinforced-Learning" usually learns from grass root stage and sometimes it also lacks the competence to prevent cyber threats or attacks at the early stage.

Based on their research it does not mention about how smart city command center leverages these deep learning algorithms to prevent the IOT attack and the information related to predictive maintenance of these devices is not yet addressed and how well we need to secure the edge devices firmware and it does not talk about threat modeling needs for command center to prevent DDOS attacks. We analyzed further papers below to understand.

As mentioned by Jyoti Deogirikar, Amarsinh Vidhate in "Security attacks in IoT: A survey paper" of IEEE [5], various types of security attacks in IOT (Network attack, Firmware attack, Botnet attack, Physical Device attack, embedded software attack, Injection attack) and their survey details about comparison of different IOT attacks (Inserting Malicious nodes, Routing, Malicious code, and side channel information). They stressed the importance of detection chances and its damage level while analyzing different attacks.

In the "Deep Learning and Big Data Technologies for IoT Security" [6] paper shared by "Mohamed Ahzam Amanullaha, Riyaz Ahamed Ariyaluran Habeebb, Fariza Hanum Nasaruddinc, Abdullah Ganid, Ejaz Ahmede, Abdul Salam Mohamed Nainarf, Nazihah Md Akimb, Muhammad Imrang" are deliberate more about the unification "Big Data", "Deep Learning" and "IOT Security". Following the key observations which helps us to continue the research more on how AI plays important role for "IOT" and "Edge Network" for Smart City command center.

1. They have picked key uses case for IoT security and its usage of deep learning and big data technology, we correlated the use cases with Smart City command centers.

| Use case | IOT Security Facet | Applicable to our Research on Smart City Command Centre area |
|---|---|---|
| SirenJack | Intrusion Detection | Using Anomaly detection with help of Neural network model (e.g., RBM (Restricted Boltzmann Machine) are used for Smart Grid data network to identify the Intrusion detection (using RNN (Recurrent Neural Network model) at Command Center. |
| DDOS Attack - Turning Up the Freeze [6] | DDOS (Distributed Denial of Service) | This use case is vital for Smart Buildings which are managed by government or private groups that are monitored centrally from command center. If we see UDP data flooding or broadcast attack or barrage attack from IOT Network (Gateway or direct devices or edge network) from these buildings, then it can be easily detected and prevented at command center using deep learning model which is built using DRNN (Dense Random Neural Network) |

| Attack on Dyn (a leading DNS provider) | DDOS (Distributed Denial of Service) | This is another area where plenty of DNS servers are maintained for Smart Parking, Smart Lights, Smart Building, and e-Governance related application, usually these DNS servers will interact with lot of Mobile applications and Wearable devices which might fall into "DDOS" attack. Command center team can collect these DNS Logs data as mentioned in their paper and it can be prevented by looking anomalous data behavior within the respective DNS server. We planned to leverage Discriminative Restricted Boltzmann Machines [7] (DBRM) model to predict the DDOS attacks through DNS Log data. |
|---|---|---|

**Table 1: Use cases matching with Smart City Command Center**

2. We are focusing more on the Network layer which is critical for smart city command center team to monitor IOT Edge, Gateway and IOT Edge to cloud network communications however the IOT Application security area is not limited to analyze the data available across networks to prevent host intrusion, Botnet attack detection and network intrusion using different deep learning models recommended in above Table 1.

In this "Integrated Command and Control Center Maturity Assessment Framework and Toolkit [8] document shared by Ministry of Housing & Urban Affairs India" has got how integrated command center collects different IOT sensor data and the role of "Command and Control Centre" focuses mainly on the following functions.

1. User interface and Visualization to generate the reports and dashboard with integrated smart city data.
2. Device Control & Monitoring – Remote monitoring and controlling of devices and event processing of large IOT data, device diagnostics.
3. Data Management where it takes of complete data operations (data transformation, metering, and control) and service management emphasis more on service management, API Management and Policy management
4. User management stress the importance of User life cycle management, Access, and Authorization management

In their paper they mentioned about security of Integrated Command and Control Centre and what should be done to avoid cyber threats and they have elaborated their Cybersecurity framework and security by design and Security during operations. It uses various security policy framework recommended by Government of India and IT Act however their recommendation did not emphasize the role of "Artificial Intelligence" or "Deep Learning" specific model to prevent or predict such cyber threats or attacks. In this paper we will try overcoming that limitation by analyzing different AI model could be used at command-and-control center.

### III. FUNCTIONAL NEEDS

Based on the above findings & limitations from different research paper that we need to focus more deeper on the IOT Security analytics for Smart City command Centre.

**Problem Statement:** We are going to discuss key IOT Security use cases of AI in smart city command centers. The following use cases are highlighted in Table 2. We will be explaining the detailed approaches in Section V.

| Use Case# | Use Case Name |
|---|---|
| 1 | Identification of anomalous traffic surges between IOT Edge and IOT Devices |
| 2 | Real-time intrusion Detection at IOT Gateway using Event Log data |

**Table2: Functional Use cases (Identified)**

### IV. SMART CITY COMMAND CENTER AND IOT NETWORK

The IOT network in Smart city are classified into three categories. The following are.

  a) Device to City Cloud communication
  b) Device to Gateway communication
  c) Device to Edge/Fog communication

a) Device to City Cloud communication: In Device to City cloud (Private) communication, the IOT Sensors (Temperature, Humidity, Pressure, Air Quality Index, Fire, Water Leaks, Power shutdown, Building HVAC systems), Physical devices (Drones to monitor citizen security, Drones for watching public accidents or traffic congestions), Wearables (Smart Watches, Wearables for Citizens who wants to leverage city services directly from City applications (e.g. City Hospital Network, Recommended Monuments and places, Recommended Parks and Parking availability for their two or four wheelers) ) are directly connected to City's own cloud or Public cloud where city allowed to communicate these devices or sensors to share the data to them. The smart city command and control team collects these directly from their cloud network or IOT Microcontroller. The data coming from actual devices are constantly monitored from both the directions (directly from microcontroller logs as well City Cloud data coming from different IOT devices) as mentioned in below diagram (Figure1)

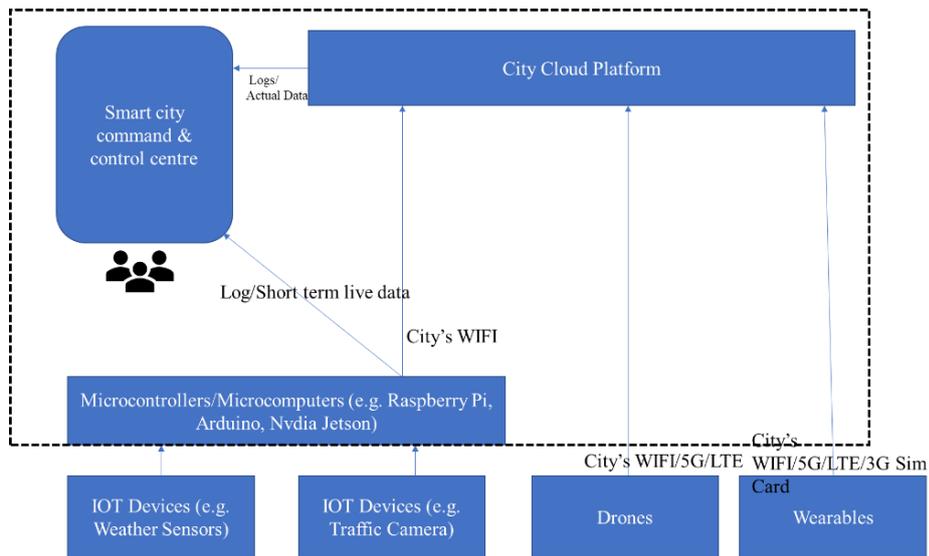

**Figure 1: Device to Cloud communication using VPN network.**

Here Control Centre team will closely monitor the data coming from micro controller through IP network which are directly connected with in the Virtual Private Network of City.

b) Device to Gateway Communication: IoT gateway in Smart City network helps with array of communication protocols between different sensors or devices that are part of City network (e.g., Smart Building, Smart Parking Space, Smart Light poles, Public WIFI Routers, Weather station sensor, Drones, etc..). It also keeps track the IOT sensors, devices and it helps to do protocol translation, data processing, data filtering and securing the devices. Below diagram (Figure 2) will show the data flow different IOT Sensors, Devices to IOT Gateway.

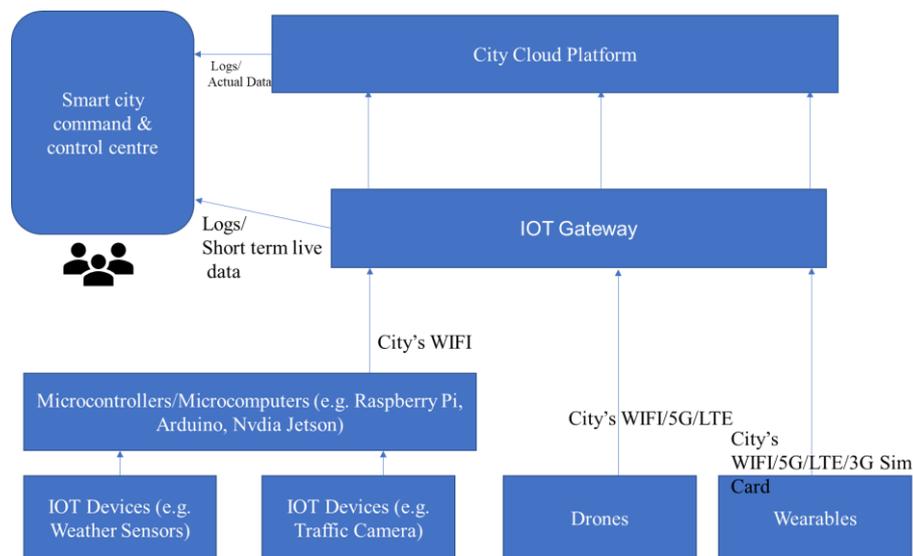

**Figure 2: Device to Gateway communication**

In this model, command and control Centre will collect data IOT Gateway devices and IOT Cloud. The monitoring and troubleshooting are feasible directly connecting to IOT Gateway IP Address.

c) Device to Edge communication: In this model, Edge network plays an important role for monitoring and controlling the device and it adds intelligence to the device to predict any anomalies or traffic surge from the sensors or devices that are intricately connected. The aggregated data from edge network will be transmitted to IOT Gateway for better translation, filtering and further aggregation required at gateway level. The gateway will communicate directly with the city's cloud. Similarly, the IOT Edge can also communicate with cloud if required. The command & control Centre team closely monitor the IOT security attacks across gateway, IOT Edge network and directly from Cloud.

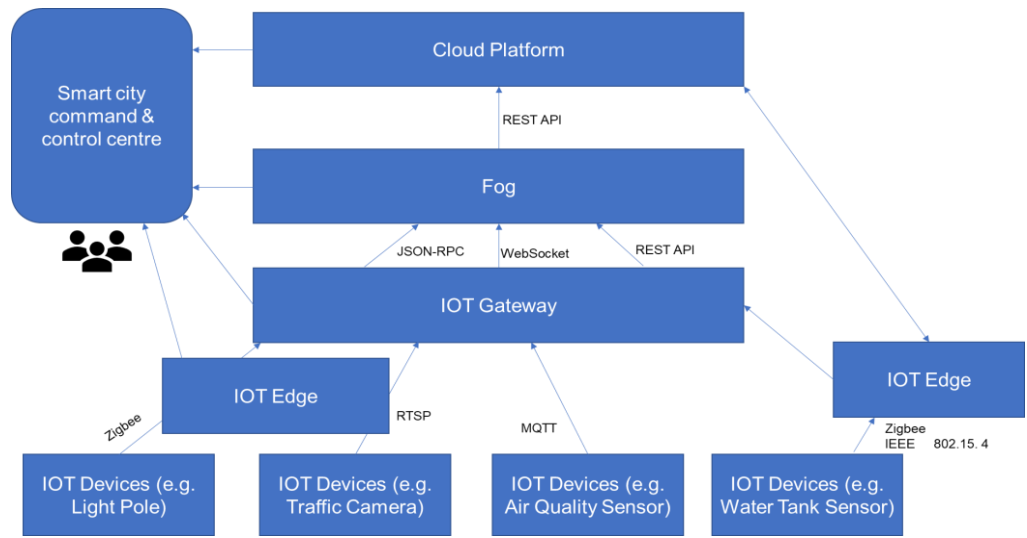

**Figure 3: Device to Edge communication**

The command & control Centre team will take care of following types of data across different smart city application, this table 3 explains the level of data security analysis required for each application.

| Functional Area of Smart City | Type of Data | Potential AI Model for IOT Network/Data/Applications |
|---|---|---|
| Smart Street Light | Light Sensing Photocells, Geo Location, Solar Panel Details, Edge Node and Cloud server details. | Predictive Maintenance and traffic surge |
| Smart Parking & Public Space management | Parking Location details, Geo Location, Payment Gateway, Vehicle Details (Car, Van, Truck, etc..), Vehicle Owner details, Building and parking lot details | Intrusion Detection<br>DDOS attacks<br>Forecasting of Network Traffic Surge |
| Smart Water Management | Water Meter Sensors, Water level sensor at different tanks, sumps and dams and lakes, water allocation data. | Intrusion Detection<br>DDOS attacks<br>Predictive Maintenance |
| Public Safety | CCTV camera, street camera, body camera for Security officials, Drone System for Surveillance. | Predictive Maintenance<br>Forecasting of Network Traffic Surge<br>DNS Spoofing or DNS Hijacking |
| Smart Meters & Grid | Grid Details, Meter Details and Geo Location, Smart Plugs, Smart Switches and Panel Board which give physical locations and Smart Transformer and Distribution Boxes. | Predictive Maintenance<br>Forecasting of Network Traffic Surge<br>GPS Spoofing |
| Smart e-Governance | Inter Department applications and its data.<br>Citizen's data | DNS Spoofing or DNS Hijacking<br>DDOS attacks |
| Smart Transport & Traffic Management | Traffic Signal Sensor data, Traffic Cameras, Smart Bus, Government Trucks and Traffic Police Vehicle Geo Location details, Traffic Control Room data. | DNS Spoofing or DNS Hijacking<br>DDOS attacks<br>Predictive Maintenance<br>GPS Spoofing |
| Smart Healthcare Management | Wearables provided by Government hospitals, | DNS Spoofing or DNS Hijacking<br>DDOS attacks |

|  | Hospital details, ambulance tracking data and ambulance geo location details, hospital services data, doctors network. |  |
|---|---|---|
| Smart Tourism & Crowd Management | Tourist arrival and departure data, Wearables recorded by Tourists or provided by Government for better tracking, Public Monument and Park details and CCTV Camera to track the crowd density and occupancy level at different places. | DDOS attacks<br>Intrusion Detection |

<div align="center">Table 3: Smart City Application's AI Model in IOT network</div>

The detailed logs for different layers of smart city network (e.g., Device/Edge, Fog, Cloud and Web Application) are usually collected in Enterprise Log Software like (Splunk, LogDNA, Elastic Stack, Sumo Logic, Fluentd, Loggly). These log database or applications will help on analyzing the logs through its catalogue of Indexes and relates real-time data in a searchable logs of event store from which it can generate meaningful insights (e.g., reports and dashboards).

The below diagram shows how IOT application and network logs can be collected and analyzed by Command & Control Centre team. It will act as one of the important sources for building AI based model to predict the following.

a. Network Traffic flow analysis
b. Network Intrusion detection & DDOS attack
c. TCP/UDP packet analysis for (Behavioral analytics)
d. Prediction of lateral movement using Domain Controller Logs
e. Malware analysis and user level calls
f. DNS Spoofing analysis

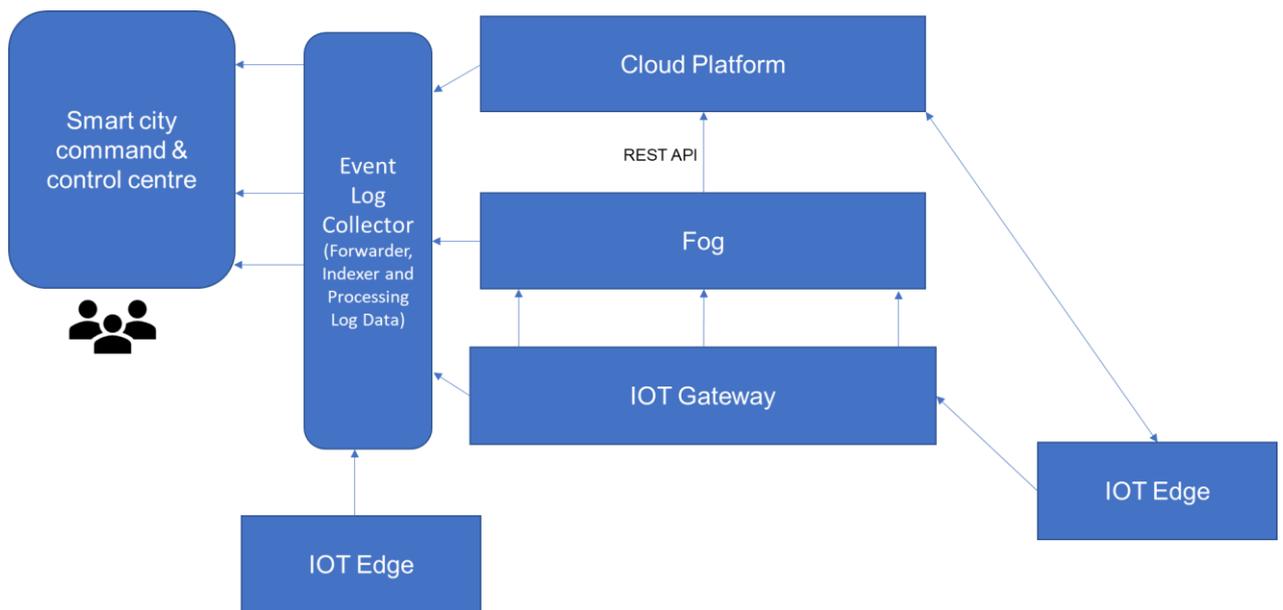

Figure 4: IOT Data Collect through Event Log Collector

| Layer | Types of Log Data | Purpose |
|---|---|---|
| Sensors and Actuator Layer | Event Log from Devices | It used to detect the heartbeat of the sensor or device and will failover easily when one of the devices is down or inactive. |
| Network Layer | IOT Edge Server and Gateway Way Server Event Log<br>Network Packet Log | It used to analyze the health of IOT Edge, and Gateway server and the network packet log helps to detect the pattern or unexpected surge in the network traffic using packet log. |
| Application Layer (Cloud) | Application Event Logs<br>Cloud Server Logs | It helps to detect the malware addresses and helps to support from known allowed and unpublished addresses. It helps to block traffic and monitoring suspected malware addresses. |

Following are the common Artificial model used in the context of Cyber Security, these AI models are agnostic to use at different network layers of IOT from the smart city context.

   a. Classification: It is used to types of similar attacks such as different pieces of malware be associated to the similar group which has related characteristics and behavior even if their signatures are different.
   b. Clustering: Cluster and Understand the different types of attacks in a TCP or UDP packets, Once the network (e.g., TCP/UDP) packets are classified into Normal and Attack using classification models, it is essential to understand the behavior relationship between each attack type using Clustering algorithms. Examples are Backdoors, DoS (Denial of Services), Exploits, Flood and Worms, etc.
   c. Predictive Models: Using Deep Learning (Neural Networks), its feasible to identify the threat as they occur. The dynamic approach will allow multiple algorithms to optimize the learning (i.e., using Auto Learning or Self Learning) capabilities routinely.
   d. User Behavior Analysis: Identification of attack's attempt at compromising or fraud by malicious users at the movement when they occur.

## V. USE CASES & HIGH-LEVEL APPROACH

**Use Case#1: Identification of anomalous traffic surges between IOT Edge and IOT Devices:**

In this use case the data coming from different IOT devices, sensors used across the smart city network. The data communication between IOT Edge and devices are happening through its native protocol [9]. In the below example we have shown the data flow from IOT Edge network and IOT Gateway network to Smart city control Centre.

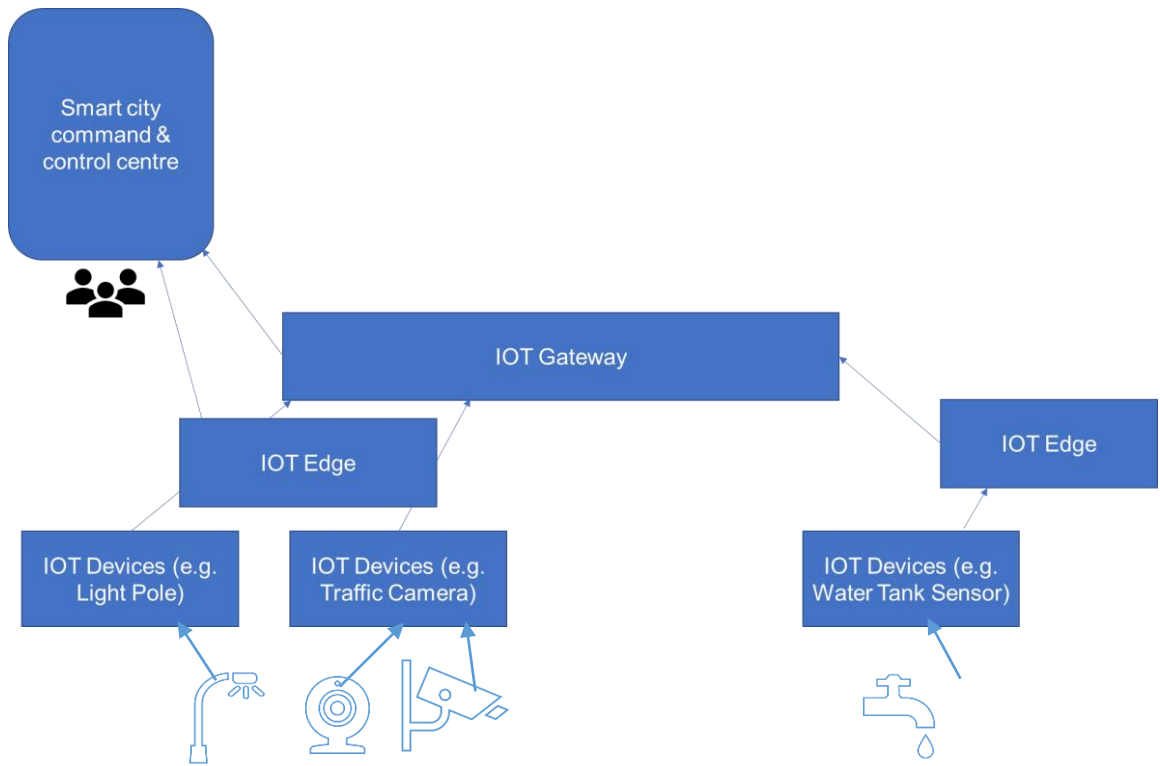

**Figure 5: Use Case1: Data Flow from IOT Network to Command-and-Control Center**

Illustration of technologies used for above use case mentioned in the following table 4.

| Device/Edge/Sensor Gateway | Technology/Platform | Details | Probable IOT Security Attack |
|---|---|---|---|
| Device (e.g., Smart Street Light [10]) - Light dependent resistor | Communication Protocol [11]: IEEE 802.15.4 ZigBee (Mid-Range RF Protocol to handle communication between Device to IOT Gateway in the IOT Edge network. | Photocells identify if streetlight [12] is needed to be turn on When the light is too low, the sensor communicates with the computing device in streetlight to activate the electricity from solar panel which sits on top of the lamp post. When the photocell identifies light, the sensor inside the streetlight post or device will deactivate the light based on the photocell values. | Intruders potentially overflow the IOT Edge Switch Buffer limit. The Edge Network Switch send the alert to command & control center that its reached buffer overflow limit. It stops communicating with Gateway which does not give the streetlight status to command & control center. |
| Smart Cameras used in Traffic Poles/Public Safety Cameras [13]. | Communication Protocol: IEEE 802.11.x Wi-Fi (Mid-Range) Protocol with 2.4GHZ/5GHz frequency to handle communication between CCTV Camera [15] | Closed-Circuit Television (CCTV) Example PTZ (Pan Tilt Zoom) and Fix box cameras to capture vehicle crossing each signal across the city. | DoS (Denial of Service) is a type of security attack [14] that seeks in ingesting Wi-Fi Bandwidth which are required for Cameras. Due to DOS attack, the camera |
| | to Edge Router in the Edge Network. Also, it will be able to manage video flow from IP cameras as RTSP streams [13] | Face Recognition cameras installed at sensitive areas will help identify miscreants or offenders. | content is not visible or even it will not send necessary packets to IOT Edge Gateway. Attackers use UDP Traffic Surge or Flooding attack where Various UDP datagrams are spawned typically by a bot. It defeats the purpose of monitoring stops sharing the necessary details to command-and-control Centre team who are supposed to monitor for any specific public safety incidents (e.g., theft, vandalism). |
| Smart Water Tank Communication [16] | Communication Protocol: The water detection level sensor uses IEEE 802.15 LoRa (Low Power Wide Area Network) for communication between sub tanks and main tank, it supports Long range wireless data transmission (10-15km) | It is used spotting that Water level in the water tank is above or below from specified threshold (e.g., 2 Feet or 5 Feet or 10 Feet) and It constantly examine whether the water level is up or down from the level that the sensor reports. The central water tank decides whether to stop the data or send the data based on the sensor reading, the central water tank details will be constantly monitored at Command & Control Centre to fulfil if any complaint from Citizen on water leak or "No" water | Routing attack is possible as many water tanks across city relate to common LPWAN Network. Sybil Attack or Sinkhole attack are high probable where they could create numerous fake identities and simulates to be different peer-to-peer networks across the Water Taps, Water Tanks connected in the IOT Network. |

**Table 4: Use Case1: Technology, Probable IOT Security Attacks**

Here we referred the Wi-Fi Sensor Gateway and Wi-Fi Router with below configuration for above devices if it needs to connect through Wi-Fi.

| Wireless Sensor Gateway | 2.4Ghz, 100m range, physical interface, include RS232, Ethernet and Zigbee communications, capable of supporting 400+ concurrent sensor nodes |
|---|---|
| Wi-Fi Router | 2.4GHz, Wi-Fi 100m + range outdoor. |

From above use case it demands for the Identification of anomalous traffic flows between IOT Device and IOT Edge/Gateway Server. The detailed experiments and model details are covered in section VI.

**Use Case#2: Real-time intrusion Detection at IOT Gateway using Event Log data:**

In this use case, the command-and-control center's log server (e.g., Splunk [17] or Elastic Search [18] or any logging software like BMC Patrol [19], Windows or Unix server logs) which continuously receives the event log of IOT Devices (Sensors) from IOT Gateway (Gateway Devices). The sufficient data protection & intrusion prevention mechanism should be done as pre-requisite (i.e., disabling root access to IOT Gateway server and resetting default root password and ensure the IOT devices are registered in the IOT Device registry and established sophisticated firewall IOT Gateway and Edge and Cloud network). Intrusion prevention system contributes as an additional layer of protection to the system to prevent from m the reactive intrusions or threats whereas Intrusion detection system tries to detect a threat and inform the command & control Centre network administrator to take appropriate action in proactive manner. This AI based real-time analytics [20] [21] helps command and control Centre [22] team to identify the intrusion proactively and take spot decision based on the prediction results against the network event logs. The detailed experiments and model details are covered in section VI.

## VI. RESEARCH APPROACH & EXPERIMENTS

In this section we describe the procedures followed for couple of use cases mentioned above including model creation and dataset used for data preprocessing, experimental scenario, results, and explanations.

**Use Case#1 Identification of anomalous traffic surges between IOT Edge and IOT Devices:**

We approached an AI model for estimating the traffic amounts in different directions in the IOT network, we need to identify if there is sudden surge in traffic [23] in unusual directions. Identification of security threats on volume for source or destination will help us to know the attacks clearly. the below process being used to deploy the AI Model for network traffic flow prediction. Following dataset are used for experimenting the network traffic surge.

| Dataset Name | Features | Smart City Use case | Relevancy to this research |
|---|---|---|---|
| Unified Host and Network Data Set [24] | - Network Event Data (11)<br>- Host Event Data (21) | Parking Sensor, Smart Street Lights and Water Meters | Intricately links with IOT network dataset and ideal for spotting various kinds of time series surges in the data and identifying outliers, suggesting to the command & control center team, and then can be further analyzed |
| MERL (Mitsubishi Electric Research Labs) [25] | - 200 wireless motion sensors | Public Safety, Crowd Management | Identification of security threats on volume for destination (e.g., a street or public place like Zoo or Monuments) |
| Intel Lab Data [26] | - Total of fifty-four IOT Sensors installed in the Intel Berkeley Research lab and it has around eight data features | Smart Buildings | Links different sensors used with in the building to identify any DDOS attacks for specific IOT Gateway or IOT Edge server. |

**Table 5: Use Case1: Data Set analyzed**

The below diagram explains the step-by-step process of building forecasting model for IOT network traffic surge from IOT Gateway or IOT Edge server.

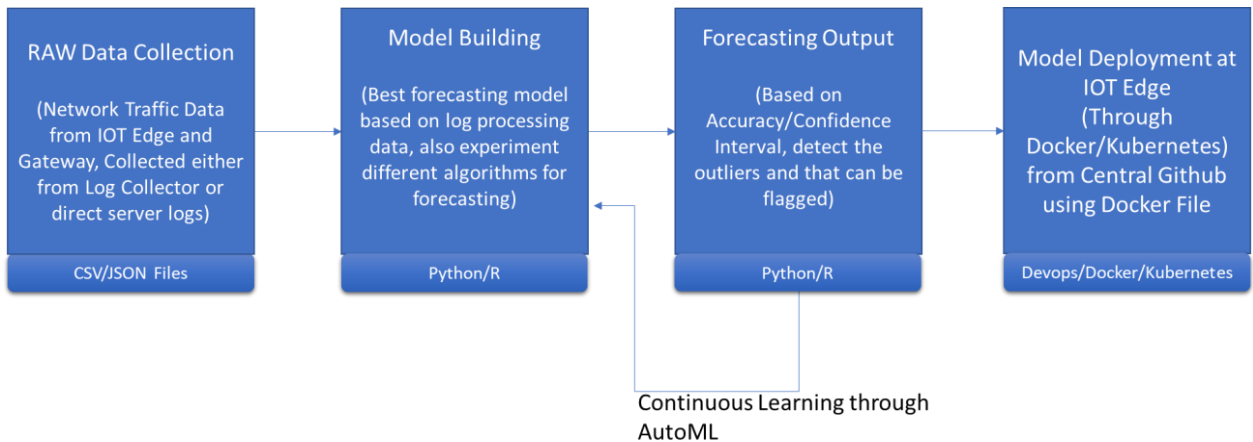

**Figure 6: Use Case1: Model Life Cycle**

RAW Data Collection: The source data from different IOT Gateway or Sensor networks will be ingested through a code-based technology (e.g., Python or R) or tool-based technologies like Azure Data Factory or AWS glue or Google Data Flow based on the cloud environment which supports the direct connectivity with IOT Gateway through REST API or other connectors like XML/JSON. The RAW data needs to be parsed and features needs to be selected for model building. Usually, we face following issues in the RAW data collection step.

a. Data Quality Errors (poor data from the identified features which could not help us to interpret any meaningful results)
b. Data skewness: It primarily refers to a non-uniform distribution in a dataset, particularly for forecasting kind of model data needs to undergo Seasonality and Stationary Testing.

Data Exploration: Post data collection, we used machine learning feature selection methods (e.g., High Correlation filter, Missing Value Ratio and Low Variance filter and) to decide the X (Time Stamp – Months/Weeks/Days/Hours) and Y axis (Forward Pockets).

Model Building: Before building the model, we need to de-duplicate the data that are collected which helps to filter and aggregate the data (e.g., milliseconds to seconds or seconds to minutes transaction roll-up) and do conversion of categorical to numerical data if applicable. We evaluated different time series AI model [27] to forecast the network surge.

| Time Series Model Aspect | Holt-Winters Model | MA | ARIMA | Linear Regression | LSTM RNN (Recurrent Neural Network) |
|---|---|---|---|---|---|
| Number of Series | 1 | 1 | 1 | 1 or more | 1 or more |
| Time Series Length | 14-200 | 12-200 | 12-200 | 14-200 | From 200 and more |
| Patterns | Seasonality and or Trend | Stationary | Stationary | Seasonality and or Trend | Any Pattern (Stationary/Seasonality) |
| MAPE (mean absolute percentage error) | Moderate | High | High | Moderate | Low |

**Table 6: Time Series Model Comparison**

We initially evaluated MA (Moving Average) and Holt Winters Model The below diagram shows the seasonality Checks done against the "Unified Host and Network Data Set" for 20 weeks data.

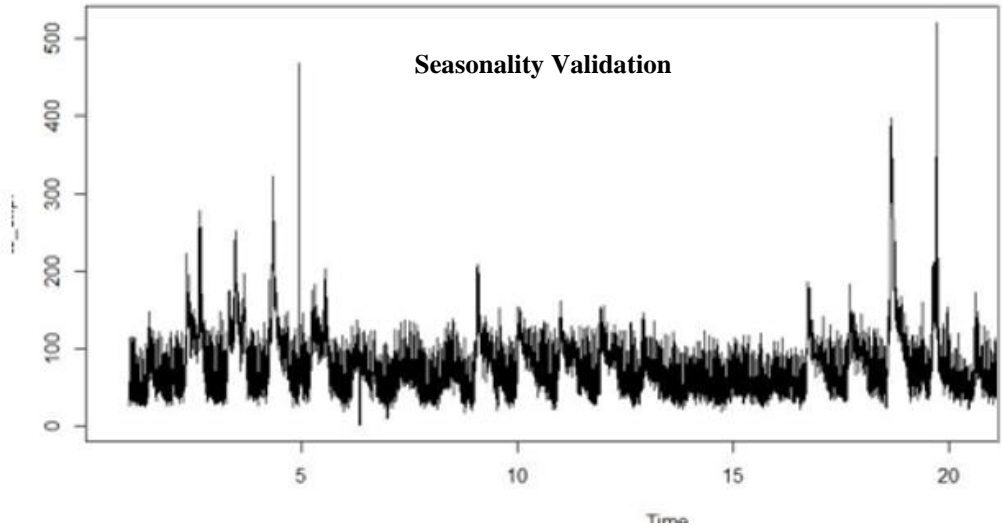

**Figure 7: Use Case1: Seasonality Variation using Holt Winter's model**

We evaluated confidence interval (CI) using the below formula.

$$X \pm Z \frac{s}{\sqrt{n}}$$

S – standard deviation

n – Number of observations or experiments

X is the mean number of TCP Packets for 4 weeks sample.

Z - Z Score value from below table 7.

|       | Z     |
|-------|-------|
| 80%   | 1.282 |
| 85%   | 1.440 |
| 90%   | 1.645 |
| 95%   | 1.960 |
| 99%   | 2.576 |
| 99.5% | 2.807 |
| 99.9% | 3.291 |

**Table 7: Z Score**

The confidence interval value is highlighted in Oval in below diagram.

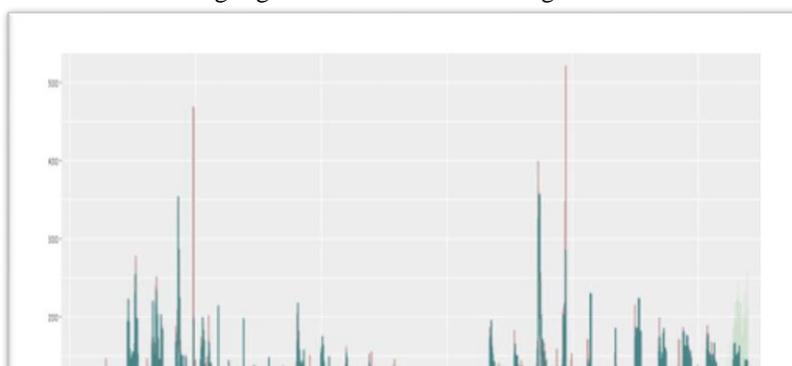

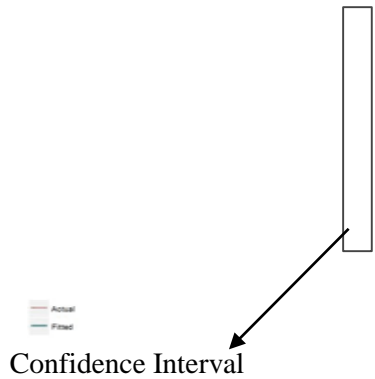

Confidence Interval

**Figure 8: Use Case1: Confidence Interval from Holt Winter's Model**

We Compared Holt-Winters and MA (Moving Average) model, Holt Winters having a better fit due to seasonal characteristic of the dataset. This is consistent with the use case as traffic data has tendency towards seasonal characteristic. We continue tested another AI Model LTSM RNN model to fine tune the confidence and increase the accuracy level.

LSTM RNN Model: We evaluated all the above models however we did a deep analysis on LSTM RNN Model [32]. Long short-term memory (LSTM) is part of recurrent neural network (RNN) architecture used mainly for Deep Learning models. LSTM networks are well-suited to building prediction model for time series data.

RNN Model: as per towardsdatascience.com science paper "Recurrent means the output at the current time step becomes the input to the next time step. At each element of the sequence, the model considers not just the current input, but what it remembers about the preceding elements."

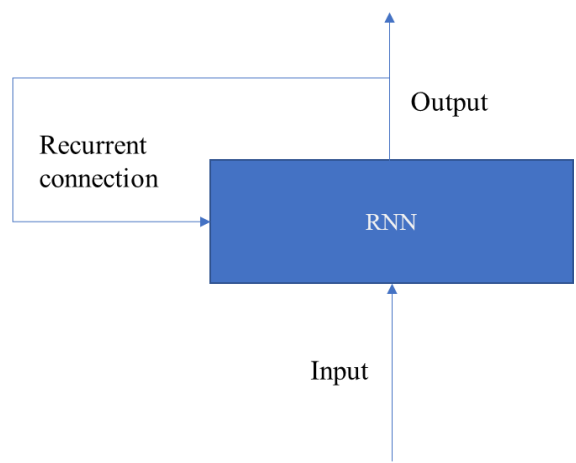

**Figure 9: RNN Model (Neural Network)**

This memory lets the network to understand long-term dependencies in a sequence that means it take up the whole perspective into account while creating a prediction model, whether that be the next packet in a network traffic data.

Model Analysis Output are shown below, the Colab screenshots are captured here.

```
[8]  # To display the top 5 rows
     df2.head(5)
```

|   | Flow ID | Timestamp | Fwd Pkt Len Mean | Fwd Seg Size Avg | Init Fwd Win Byts | Init Bwd Win Byts | Fwd Seg Size Min |
|---|---|---|---|---|---|---|---|
| 0 | 172.31.69.28-18.216.200.189-80-52169-6 | 22/02/2018 12:27:57 AM | 233.750000 | 233.750000 | -1 | 32768 | 0 |
| 1 | 172.31.69.25-18.219.193.20-80-44588-6 | 16/02/2018 11:18:14 PM | 0.000000 | 0.000000 | -1 | 225 | 0 |
| 2 | 172.31.69.25-18.219.193.20-80-43832-6 | 16/02/2018 11:23:20 PM | 114.333333 | 114.333333 | -1 | 219 | 0 |
| 3 | 172.31.69.25-18.219.193.20-80-53346-6 | 16/02/2018 11:22:41 PM | 233.750000 | 233.750000 | -1 | 211 | 0 |
| 4 | 172.31.69.28-18.218.55.126-80-57856-6 | 21/02/2018 11:49:25 PM | 233.750000 | 233.750000 | -1 | 32768 | 0 |

Time Series start and end date details are shown below.

```
df2['Fwd Pkt Len Mean'] = pd.to_numeric(df2['Fwd Pkt Len Mean'], errors='coerce')
df2 = df2.dropna(subset=['Fwd Pkt Len Mean'])

df2 = df2.loc[:, ['Timestamp', 'Fwd Pkt Len Mean']]
df2.sort_values('Fwd Pkt Len Mean', inplace=True, ascending=True)
df2 = df2.reset_index(drop=True)

print('Number of rows and columns after removing missing values:', df2.shape)
print('The time series starts from: ', df2['Timestamp'].min())
print('The time series ends on: ', df2['Timestamp'].max())
```

```
Number of rows and columns after removing missing values: (499563, 2)
The time series starts from:  03/07/2017 05:25:58 PM
The time series ends on:  22/02/2018 12:35:54 AM
```

LTSM RNN Model:

define the shape of the input dataset: Since we are only using one feature of Fwd Pkt Len Avg.

```
import tensorflow as tf
from tensorflow import keras
from tensorflow.keras import layers
from tensorflow.keras.utils import Sequence
ts_inputs = tf.keras.Input(shape=(num_timesteps, 1))

# units=10 -> The cell and hidden states will be of dimension 10.
#             The number of parameters that need to be trained = 4*units*(units+2)
x = layers.LSTM(units=10)(ts_inputs)
x = layers.Dropout(0.2)(x)
outputs = layers.Dense(1, activation='linear')(x)
model = tf.keras.Model(inputs=ts_inputs, outputs=outputs)
```

```python
model.compile(optimizer=tf.keras.optimizers.SGD(learning_rate=0.01),
              loss=tf.keras.losses.MeanSquaredError(),
              metrics=['mse'])
```

Current date and time :
2021-05-17 14:52:05

[57] `model.summary()`

```
Model: "model_2"
_________________________________________________________________
Layer (type)                 Output Shape              Param #
=================================================================
input_4 (InputLayer)         [(None, 1008, 1)]         0
_________________________________________________________________
lstm_2 (LSTM)                (None, 10)                480
_________________________________________________________________
dropout_2 (Dropout)          (None, 10)                0
_________________________________________________________________
dense_2 (Dense)              (None, 1)                 11
=================================================================
Total params: 491
Trainable params: 491
Non-trainable params: 0
_________________________________________________________________
```

```python
BATCH_SIZE = 128
NUM_EPOCHS = 1
NUM_CHUNKS = tss.num_chunks()

for epoch in range(NUM_EPOCHS):
    print('epoch #{}'.format(epoch))
    for i in range(NUM_CHUNKS):
        X, y = tss.get_chunk(i)

        model.fit(x=X, y=y, batch_size=BATCH_SIZE)

    # shuffle the chunks so they're not in the same order next time around.
    tss.shuffle_chunks()
```

```
epoch #0
100/100 [==============================] - 25s 233ms/step - loss: 1.0767e-04 - mse: 1.0767e-04
100/100 [==============================] - 23s 229ms/step - loss: 3.4526e-05 - mse: 3.4526e-05
CPU times: user 1min 21s, sys: 3.33 s, total: 1min 25s
Wall time: 48.6 s
```

The test dataset using LSTM shows MSE (Mean Squared Error) as "0.0000034526". While the baseline model has MSE of "0.00003428". The current LSTM model performs better than the previous (baseline) model.

How this model can be deployed at Edge server and continuously learn: The aforementioned code is developed using python and it can be compiled and built it in a docker and it can be exported to dockerhub as a "packaged" container which can be reused for further deployments, if we install a docker in the IOT edge server which could able to pull the docker image from docker hub and able to run independently or it could be orchestrated using Kubernetes (orchestation tools) using minkube using kubectl command at the IOT Edge.

How Command and Control team get benefited: if this model proven for set of IOT Gateway or IOT Edge servers, it can be auto deployed using "DevOps [28]" tools like Jenkin/GitHub and SonarQube. Also using these forecasted network surge, they could clearly understand which IOT Gateway is potentially going to be weak from the attack and they can ensure by taking actions (e.g., firmware update, password reset, adding additional firewall, protecting the devices using encrypted protocol)

**Use Case#2: Real-time intrusion Detection at IOT Gateway using Event Log data**

The below diagram explains the step-by-step process of building real-time log analytics detect the Host based Intrusion using Event Logs data.

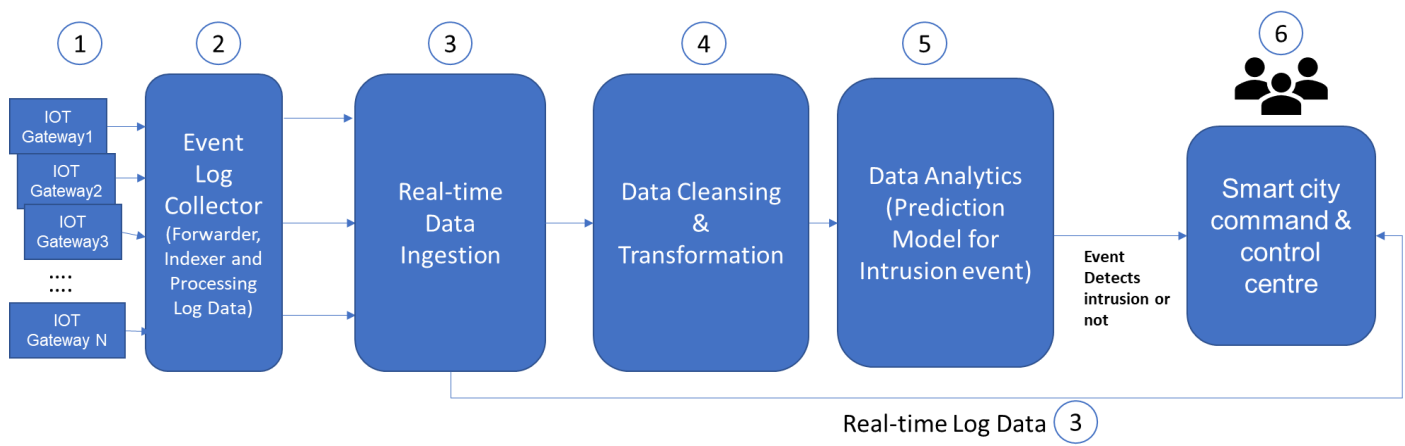

Figure 10: Use Case2: Model Lifecycle

| Step# | Activities | Tools/Technologies Referred [29][30][31][32][33] | Remarks |
|---|---|---|---|
| 1 | **IOT Gateway:** Smart City will have different IOT Gateway server in different locations to support the data coming different IOT devices (e.g., smart streetlights, smart parking sensors, building alarms, CCTV camera event logs, smart wearables) and public WIFI Router | Splunk Agent or API /Kepware/IOT Gateway Devices (Hardware) | Here we referred Kepware as IOT Gateway software which translates the protocol and takes care of translation, aggregation and sending to the Gateway Server. |
| 2 | **Event Log Collector:** collects all the different logs through log syncing mechanism (e.g., syncing from client to server like BMC Patrol or Splunk) and it will be | Splunk/Elastic Search with Active MQ or Kafka (Message Broker) | This supports streaming of logs from different IOT gateway. |
| | kept in the log collector software storage or central data storage like data lake. | | |
| 3 | **Data Ingestion:** ingestion of log stream accepts the inbound input and performs minor transformations and channels data through the real-time prediction's module.<br>persisted the output to Unstructured Storage like Data Lake (e.g., ADLS, Google Cloud Storage or AWS S3 Bucket)<br>**Real-time Dashboard: Alerts & Notification:**<br>The real-time data will go to smart city command Centre team to monitor the logs in parallel and they can build alerts and dashboards to see if there are any abnormal events or unusual IP addresses. | AWS Kinesis/Google Clod Data Flow/Azure Event Hub/Azure Data Factory/Custom code written in Python or Java.<br><br>We can build Real-time dashboards either through the log collector tool like Splunk or open-source tool like Grafana/Kibana/Banana based on the Log collection software (e.g., Elastic Search, Solr, Fluentd) | |
| 4 | **Data Cleansing & Transformation:**<br>- Unwanted will be cleansed from the Logs dataset.<br>- Symbolization (Vectorization/Encoding) will be done on the cleansed dataset.<br>- Aggregation of data (roll-up to second if the data comes from milliseconds) | Pre-Built tools from Log collector tool (e.g., Splunk) or Custom code to cleanse and transform the data through Python/R/Matlab. | |
| 5 | Data Model:<br>Supervised learning technique for identifying various classification methods to classify the threats. Identifying important feature and | Deep Learning Tools:<br>Apache Spark MLlib/Tensor Flow/Pytorch/Keras | Example: The corner classification algorithm (CC4) used for notorious intrusion attacks |
| | classification into various buckets so that a new threat could be identified. | | |
| 6 | **Real-time Prediction Output:** It will be merged with the real-time dashboard built in step 3 and it will indicate the command & control Centre team to predict the Intrusion of any specific IOT Gateway network (i.e., known packets or unknown packets or Attack packets) | Same as Step 3 | |

**Table 9: Use Case2: Detailed Steps**

Though we could not experiment this use case due to non-availability of "Real-time" data however it could be tried with a IOT Event Data simulator once it is available.

## VII. DEPLOYMENT CONSIDERATIONS

Following are the key deployment considerations for deploying the model for smart city's command-and-control center team [34][35][36][37] to get the real benefit of AI from the IOT data that are going to be collected from different IOT network or application or devices.

   a) Capability assessment for IOT Edge Servers in Smart City network based on its compute power (e.g., RAM, CPU or GPU) and whether it will be able to hold any AI model to support threat detection or prediction.
   b) Prioritize the IOT network which collects sensitive data or privacy data where it flows from a particular Smart City application (e.g., CCTV Camera, Wearable or Payment Channel which hold Citizen's sensitive and confidential information). Based on that list of IOT Gateway or IOT Edge or Devices, we need to deploy the AI model (either real-time or near real-time).
   c) While we choose the Deep Learning model, we need to consider the following parameter from deployment perspective even though they give high accuracy with test data, sometimes it may run out of resources during continuous learning at IOT Edge.
      1. Model Size (Number of Layers)
      2. Parameter Size (100,000,000+)
      3. Time and Cost (Upto Weeks and sometime up to months)
      4. Hardware (Graphical Processor Unit)
      5. Framework – Distributed.
   d) As the volume of IOT network and number of devices are significantly high for Smart City kind of applications which needs more robust method of deploying the model automatically.
   e) The network complexity, latency, and threshold data of IOT network should be collected before we target the model for a specific set of IOT Edge or Gateway servers.

- f) City command & control center should already have Intrusion Detection System apart from this model and it should help to monitor based on the intrusion specific rules (e.g., port range, firewall rules, IP network ranges, Device Registry, Router and Switch metadata, Firmware providers and details)
- g) We should eliminate hard coded passwords of Wi-Fi Routers, Gateway Server or know IOT Devices based on the provider details. Root Accounts for all logins should be disabled for vendors who keep update their software remotely, we should supply different user account for them.
- h) Keep Antivirus software and virus definitions and firewall rules up to date.
- i) Enablement of Device or IOT Edge or IOT Gateway logs centrally to monitor the data transmission for better troubleshooting and do root cause analysis. This could be achieved by installing "Log" agents which synchronize the data at the "Central" Log server (e.g., Splunk or BMC Patrol or Nagios)

## VIII. SIGNIFICANCE OF THIS RESEARCH & RECOMMENDATIONS

The population of IoT devices are significantly growing for Smart City platforms, IOT devices are exposed to Security outbreaks such as Intrusion, Flooding, Denial of Services, etc. Also, IOT devices are going to produce large dataset (structured and unstructured (e.g., Images)) which needs "AI" to go along with IOT network to help predict, detect, forecast the network attacks seamlessly from Smart City command & control center. Its time to harmonize the power of "AI" (Artificial Intelligence) at the IOT to protect and safeguard from Cyber-attacks. Due to computing power limitations, we trained the model only for selected use cases with limited number of IOT Devices. The above models can be further optimized and deployed on to IOT Edge or Gateway automatically with niche tools in the market.